\documentclass[twocolumn]{aastex61}

\usepackage{hyperref}

\begin{document}

\title{Spitzer Matching survey of the UltraVISTA ultra-deep Stripes (SMUVS):  \\
Full-mission IRAC Mosaics and Catalogs }


\author[0000-0002-3993-0745]{M.\,L.\,N.\,Ashby}
\affil{Harvard-Smithsonian Center for Astrophysics, 
Optical and Infrared Astronomy Division, 
60 Garden St., MS-66,
Cambridge, MA 02138, USA \\ {\em mashby@cfa.harvard.edu}}

\author{Karina I.\,Caputi}
\affil{Kapteyn Astronomical Institute, University of Groningen, 
P.O. Box 800, 9700 AV Groningen, The Netherlands}

\author{Will Cowley}
\affil{Kapteyn Astronomical Institute, University of Groningen, 
P.O. Box 800, 9700 AV Groningen, The Netherlands}

\author{Smaran Deshmukh}
\affil{Kapteyn Astronomical Institute, University of Groningen, 
P.O. Box 800, 9700 AV Groningen, The Netherlands}

\author[0000-0002-1404-5950]{James S.\,Dunlop}
\affil{Institute for Astronomy, University of Edinburgh, 
Royal Observatory, 
Blackford Hill,  Edinburgh EH9 3HJ, UK}

\author[0000-0002-2281-2785]{Bo Milvang-Jensen}
\affil{Dark Cosmology Centre, Niels Bohr Institute, 
University of Copenhagen, Juliane Maries Vej 30, 2100 Copenhagen {\O}, Denmark}
\affil{The Cosmic Dawn Centre, Niels Bohr Institute, 
University of Copenhagen, Juliane Maries Vej 30, DK-2100, Copenhagen {\O}, Denmark}

\author[0000-0002-8149-8298]{Johan P.\,U.\,Fynbo}
\affil{The Cosmic Dawn Centre, Niels Bohr Institute, 
University of Copenhagen, Juliane Maries Vej 30, DK-2100, Copenhagen {\O}, Denmark}
\affil{Dark Cosmology Centre, Niels Bohr Institute, 
University of Copenhagen, Juliane Maries Vej 30, 2100 Copenhagen {\O}, Denmark}

\author[0000-0002-9330-9108]{Adam Muzzin}
\affil{Department of Physics and Astronomy, York University, 
4700 Keele St., Toronto, Ontario, MJ3 1P3, Canada}

\author{H.\ J.\,McCracken}
\affil{Institut d'Astrophysique de Paris, Sorbonne Universit\`{e}s, 
UPMC Univ.\ Paris 6 et CNRS, UMR 7095, 98 bis bd Arago, 75014 Paris, France}

\author[0000-0001-5891-2596]{Olivier Le F\`evre}
\affil{Laboratoire d'Astrophysique de Marseille, 
38, rue Fr\'ed\'eric Joliot-Curie 13388 Marseille cedex 13 France}

\author{Jia-Sheng Huang}
\affil{Harvard-Smithsonian Center for Astrophysics, 
Optical and Infrared Astronomy Division, 
60 Garden St., MS-66,
Cambridge, MA 02138, USA}

\author[0000-0002-1740-6360]{J.\,Zhang}
\affil{Harvard University, 
Cambridge, MA 02138, USA}


\begin{abstract}

This paper describes new deep 3.6 and 4.5\,$\mu$m imaging  
of three UltraVISTA near-infrared survey stripes within the COSMOS field.  The observations 
were carried out with {\sl Spitzer's} Infrared Array Camera (IRAC) for the {\sl Spitzer} Matching
Survey of the Ultra-VISTA Deep Stripes (SMUVS).  In this work we present our data reduction
 techniques, and document the resulting mosaics, coverage maps, and catalogs in both IRAC 
 passbands for the three easternmost UltraVISTA survey stripes, covering a combined area of about 
 0.66\,deg$^2$, of which 0.45\,deg$^2$ have at least 20\,hr integration time.  
 SMUVS reaches point-source sensitivities of about 25.0\,AB mag (0.13\,$\mu$Jy) at both 
 3.6 and 4.5\,$\mu$m with a significance of 4$\sigma$, accounting for both survey sensitivity and source
confusion.  To this limit the SMUVS catalogs contain a total of $\sim$350,000 sources, each of which
is detected significantly in at least one IRAC band.  Because of its uniform and high sensitivity, relatively
large area coverage, and the wide array of ancillary data available in COSMOS, the SMUVS survey
 will be useful for a large number of cosmological investigations.   
We make all images and catalogues described herein publicly available via the Spitzer Science Center.

\end{abstract}


\section{Introduction} \label{sec:intro}

Measurements of galaxy number density and stellar-mass evolution at high redshifts 
($z > 3$) are the foundation for a proper understanding of how galaxy buildup proceeded 
in the early Universe.  Number density and stellar mass estimates directly constrain models 
of the candidate mechanisms for galaxy growth, such as galaxy mergers 
(e.g., Hopkins et al.\ 2006; Somerville et al.\ 2008) or cold gas accretion within gas-rich 
proto-disks (e.g., Dekel et al.\ 2009).  Whatever the mechanisms might be that govern galaxy 
evolution, they must reproduce the observed 
distribution of baryons at high redshift, and connect it to the subsequent evolution of galaxies 
within dark matter haloes.

For galaxies out to redshifts $z=2-3$, stellar masses are typically derived from broadband 
photometry between the (rest-frame) 4000\AA\ break and $K$-band (e.g., Bell \& de Jong 2001), 
because observations in this interval are more sensitive to the light from the stellar populations 
that dominate the stellar mass.  Beyond $z=3$, 
however, it becomes very challenging to photometer the stellar populations that dominate the total
stellar mass because of the
high sky backgrounds at wavelengths longward of $K$.  For galaxies 
at $z>3$, one must therefore turn to mid-infrared observations from space.  This is exactly why  
$3.6-4.5$\,$\mu$m imaging with the Infrared Array Camera (IRAC; Fazio et al.\ 2004) aboard 
the {\sl Spitzer Space 
 Telescope} (Werner et al.\ 2004) is indispensable for mapping the rest-frame near-IR 
 light from distant galaxies. IRAC observations are also necessary to identify distant active galactic 
 nuclei (AGN), particularly when the nuclear activity is too obscured by dust to be detected in X rays 
 (e.g., Lacy et al.\ 2004; Stern et al.\ 2005; Caputi 2013, 2014).
 
Recent studies of massive galaxies ($M>\sim10^{11} M_\sun$) at high redshifts have revealed
 significant number-density evolution between $z = 5$ and $z = 3$, consistent with 
  much faster assembly than between $z = 2$ and $z = 0$ (Caputi et al.\ 2011; 
 Ilbert et al.\ 2013; Muzzin et al.\ 2013; Stefanon et al.\ 2015).  
 By contrast, relatively little is known about the evolution of intermediate-mass galaxies 
 ($M\sim10^{10} M_\sun$) at $z > 3$, because typical IRAC surveys are too shallow 
 to yield complete samples of these galaxies over large areas of the sky.  This is unfortunate, 
 because the relatively numerous intermediate-mass galaxies are expected to contain most
of the stellar mass of the Universe at high redshifts (e.g., Caputi et al.\ 2015). 
Identifying and characterizing complete samples of intermediate-mass galaxies 
is therefore crucial for constraining galaxy formation models.

Optically selected galaxy samples at $z > 3$ have yielded some important information about 
intermediate-mass galaxies, such as typical Lyman-break galaxies (e.g., Steidel et al.\ 2003; 
Malhotra et al.\ 2005; Shapley et al.\ 2006).  However, these samples are not fully representative 
of intermediate-mass galaxies at high redshifts because they are biased against dust-obscured 
sources, and the stellar-mass estimates of high-redshift Lyman-break galaxies are robust 
only when IRAC photometry is available.  To obtain galaxy samples that are complete in stellar mass 
at high redshifts, it is necessary to avoid dust attenuation by selecting targets in deep infrared maps,
and it helps greatly to have coextensive {\sl Spitzer}/IRAC imaging (e.g., Caputi et al.\ 2014) in order
to get multiple measurements of the redshifted stellar continua.

In this contribution, we describe a new IRAC survey designed to provide deep rest-frame
optical/near-IR imaging over a large area of the sky for which deep ground-based imaging is available.
This survey, the {\sl Spitzer} Matching survey of the UltraVISTA ultra-deep Stripes (SMUVS),  
covers three ultra-deep stripes of the UltraVISTA survey (McCracken et al.\ 2012) within the
COSMOS field with extremely sensitive semi-continuous imaging in both operating IRAC bands.  
SMUVS is intended to provide the community with the best prospects to build upon present 
knowledge of galaxy evolution beyond $z=3$.  Figure~1 illustrates the relationship of SMUVS 
to other extragalactic surveys carried out with {\sl Spitzer}.

\begin{figure*}[ht!]
\figurenum{1}
\epsscale{1.15}
\plotone{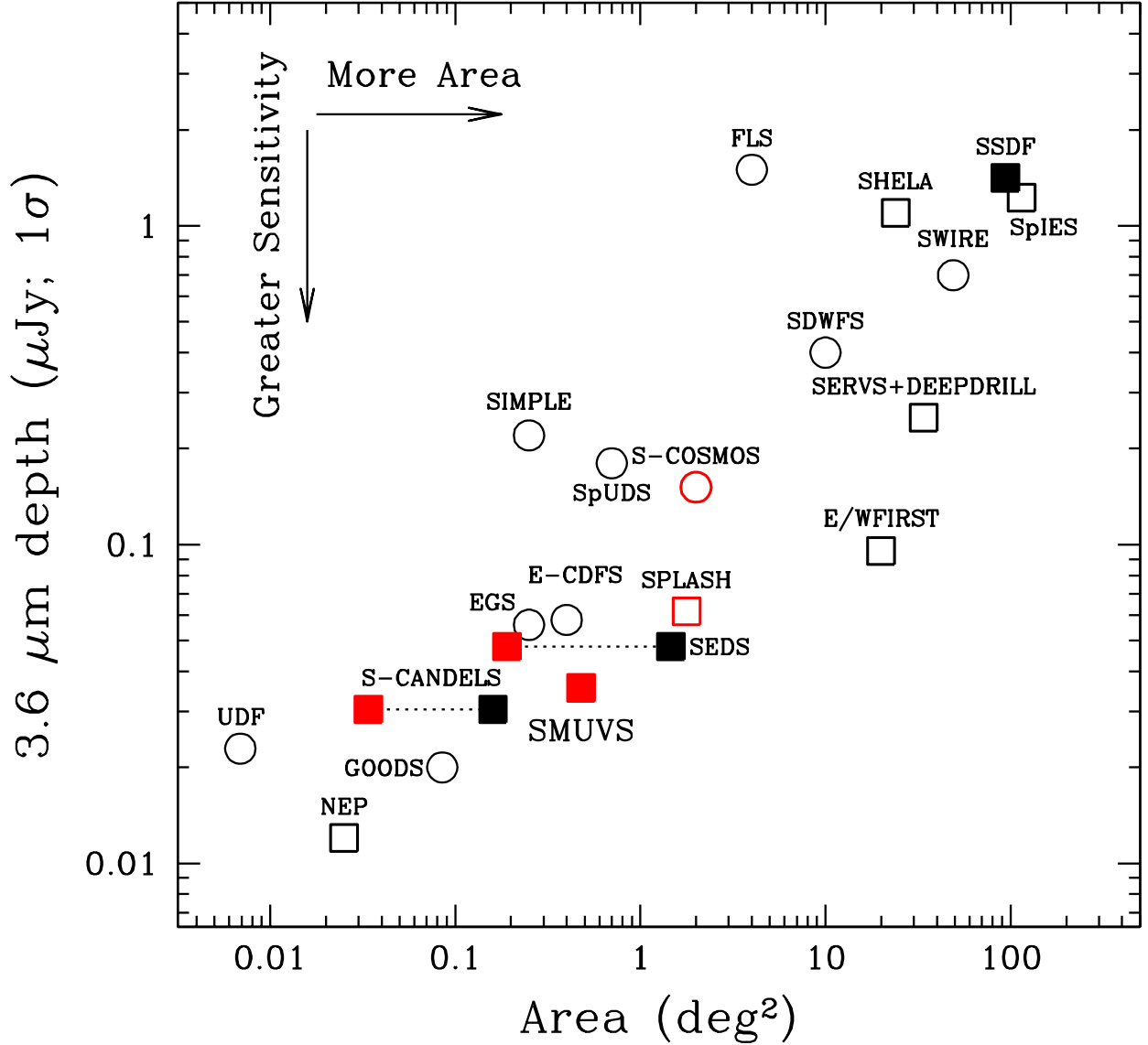}
\caption{SMUVS in relation to other completed {\it Spitzer}/IRAC extragalactic surveys.  
Circles and squares indicate the 3.6\,$\mu$m $1\sigma$ point-source sensitivities for surveys 
executed during the cryogenic and warm mission phases, respectively.  Solid squares 
indicate sensitivities calculated with simulations; the other sensitivities shown are either taken 
from the literature or from the online calculator SENS-PET under low-background conditions.
All red symbols indicate IRAC surveys carried out 
within the COSMOS field specifically, including 
S-COSMOS (Scoville et al.\ 2007), 
SPLASH (Steinhardt et al.\ 2014), 
SEDS (Ashby et al.\ 2013a),
and S-CANDELS (Ashby et al.\ 2015), in addition to SMUVS.  
SEDS and S-CANDELS are multi-field surveys for which the total and the COSMOS-specific portions 
are indicated with connected black and red squares.  Also shown are 
the First Look Survey (FLS; Fang et al.\ 2004),
{\sl Spitzer}-SPT Deep Field (SSDF; Ashby et al.\ 2013b),
{\sl Spitzer}-IRAC Equatorial Survey (SpIES; Timlin et al.\ 2016),
{\sl Spitzer}-HETDEX Exploratory Large-area Survey (SHELA; Papovich et al.\ 2016),
{\sl Spitzer} Wide-area Infrared Extragalactic Survey (SWIRE; Lonsdale et al.\ 2003, 2004),
{\sl Spitzer} Deep, Wide-Field Survey (SDWFS; Ashby et al.\ 2009),
{\sl Spitzer} Extragalactic Representative Volume Survey (SERVS; Mauduit et al.\ 2012) 
augmented by additional, contiguous coverage to the same depth from DEEPDRILL (M.\ Lacy, priv. comm.),
{\sl Spitzer} IRAC/MUSYC Public Legacy in E-CDFS (SIMPLE; Damen et al.\ 2011),
{\sl Spitzer} Public Legacy Survey of UKIDSS Ultra-deep Survey (SpUDS; Caputi et al.\ 2011),
{\sl Euclid}/WFIRST {\sl Spitzer} Legacy Survey (E/WFIRST; PI Capak),
the Extended {\sl Chandra} Deep Field South (E-CDFS; Rix et al.\ 2004),
the Extended Groth Strip (EGS; Barmby et al.\ 2008),
the Ultra-deep Field (UDF; Labb\'e et al.\ 2013), 
the Great Observatories Origins Deep Survey (GOODS; Lin et al.\ 2012), 
and
the IRAC dark field at the North Ecliptic Pole (NEP; Krick et al.\ 2009; J.\ Surace, priv. comm.).
\label{fig:etendue}}
\end{figure*}

This paper is organized as follows.  Sec.\,~\ref{sec:ultravista} describes the UltraVISTA survey.  In 
Secs.\,~\ref{sec:obs} and \ref{sec:reduction} we describe the multi-epoch IRAC observations carried out
for SMUVS and other coextensive IRAC surveys, and describe how the observations were 
reduced to catalog form.  Finally, Sec.\,~\ref{sec:validation} 
describes the tests applied to the SMUVS catalogs to validate them.  

\section{The UltraVISTA Ultradeep Survey within the COSMOS Field}
\label{sec:ultravista}

The Cosmic Evolution Survey (COSMOS, Scoville et al.\ 2007) is a well-known  
 extragalactic survey field covering $\sim2$\,deg$^2$ sited strategically at 
$(\alpha, \delta) = (10^{\rm h}00^{\rm m},+02^\circ12^\prime)$ where it is accessible 
to ground-based telescopes in both the northern and southern hemispheres.  In addition to 
high-resolution imaging  
with the largest contiguous HST/ACS survey so far compiled (Koekemoer et al.\ 2007), COSMOS benefits
from extensive imaging at X-ray, optical/infrared, submillimeter, radio, and other 
wavelengths, plus an abundance of spectroscopy.  These overlapping surveys feature a combination
of high sensitivity and wide area coverage designed to sample large volumes and
thereby facilitate a better understanding of galaxy evolution without undue complications from
cosmic variance.

The pressing need for deep near-IR photometry within COSMOS motivated a large allocation of
observing time for multiband imaging in survey mode with the VIRCAM instrument (Dalton et al.\ 2006) 
at the VISTA telescope (Emerson \& Sutherland 2010).  This effort is known as UltraVISTA 
(McCracken et al.\ 2012).  UltraVISTA is the deepest of the public surveys being carried out 
with the VISTA telescope.  No other near-IR survey covers as much area as deeply as UltraVISTA.  
Specifically, UltraVISTA has mapped  $\sim1.8$\,deg$^2$ of COSMOS in  
$YJHK_s$, plus half that area in the $NB118$ narrowband filter at $1.19$\,$\mu$m (Milvang-Jensen et al.\ 2013). 
UltraVISTA consists of two main parts: a deep survey reaching 
$K_s = 23.7$\,AB mag (5$\sigma$) over the full area, and an ultra-deep 
survey that will reach $K_s\approx25.3$ and $H\approx25.7$\,AB mag (both 5$\sigma$) 
over four stripes covering a total of $\sim0.8$\,deg$^2$ in the final data release 
(Fig.~\ref{fig:layout1}; see also Fig.\,1 of
McCracken et al.\ 2012).   After 8 years of observations that started at the end of 2009
the primary UltraVISTA survey is now essentially complete.
The forthcoming DR4 data release will contain stacks based on re-reduced
data for the first 7 years (DR3 corresponded to the first 5 years).  
In addition, a new UltraVISTA extension program that began in 2017 April 
will enlarge the area of homogeneous ultra-deep $JHK_s$ coverage to $\sim1.8$\,deg$^2$.  

\section{IRAC Mapping of the COSMOS Field}
\label{sec:obs}

\begin{deluxetable}{l@{\hspace*{3em}}l@{\hspace*{3em}}c@{\hspace*{3em}}l}
\tabletypesize{\footnotesize}
\tablewidth{344pt}
\tablecaption{{\sl Spitzer}/IRAC Imaging Campaigns in COSMOS\label{tab:obslog}}
\tablehead{
{PID\tablenotemark{a}} & {Epoch} & {Approximate T$_{\rm INT}$} \\
&  & {(hours)}
}
\startdata
\multicolumn{3}{c}{SMUVS STRIPE 1 (10:02, +2:18) } \\
\hline
20070 & 2005 Dec 30--2006 Jan 02  & 0.3 \\
90042 & 2013 Feb 02--Mar 04       & 1.7 \\
90042 & 2013 Jul 04--Aug 07         & 1.7 \\
90042\tablenotemark{b} 
           & 2014 Feb 17--Mar 10        & 1.4 \\ 
10159 & 2014 Jul 13--Aug 19         & 0.6 \\
11016 & 2015 Feb 13--Mar 17        & 9.0 \\
11016 & 2015 Jul 21--Jul 30           & 9.0 \\
11016 & 2016 Mar 01--Mar 22         & 2.2 \\
11016 & 2016 Aug 16--Sep 03           & 15.4\ \ \\
11016 & 2017 Feb 26--Apr 04          & 4.4 \\
\hline
\multicolumn{3}{c}{SMUVS STRIPE 2 (10:00:30, +2:14) } \\
\hline
20070 & 2005 Dec 30--2006 Jan 2  & 0.3 \\
61043 & 2010 Jan 25--Feb 04          & 4.0 \\
61043 & 2010 Jun 10--Jun 28          & 4.0 \\
61043 & 2011 Jan 30--Feb 06          & 4.0 \\
80057 & 2012 Feb 04--Feb 19         & 36.0\ \  \\
80057 & 2012 Jun 26--Jul 09           & 36.0\ \  \\
11016 & 2015 Feb 24--Mar 19          & 9.0 \\
11016 & 2015 Aug 22--Aug 27           & 9.0 \\
11016 & 2016 Mar 02--Mar 21          & 6.7 \\
11016 & 2016 Jul 29--Aug 19            & 8.3 \\
11016 & 2017 Mar 02--Mar 05           & 1.5 \\ 
\hline
\multicolumn{3}{c}{SMUVS STRIPE 3 (9:59, +2:13) } \\
\hline
20070 & 2005 Dec 30--2006 Jan 2  & 0.3 \\
90042 & 2013 Feb 02--Mar 04         & 1.7 \\
90042 & 2013 Jul 04--Aug 07           & 1.7 \\
90042 & 2014 Feb 17--Mar 10         & 1.4 \\ 
10159 & 2014 Jul 13--Aug 19          & 0.6 \\
11016 & 2015 Feb 12--Mar 18         & 6.7 \\
11016 & 2015 Jul 21--Aug 07           & 6.7 \\
11016 & 2016 Mar 17--Mar 23          & 2.2 \\
11016 & 2016 Jul 29--Aug 15           & 7.0 \\
11016 & 2017 Mar 01--Apr 04            & 16.7\ \   \\
\enddata
\tablecomments{{\sl Spitzer}/IRAC observations of the three UltraVISTA stripes covered by SMUVS.  Integration times are illustrative only, due to significant variation by position within each SMUVS epoch.  
Coverage is not necessarily coextensive on successive epochs.}
\tablenotetext{a}{{\sl Spitzer} Program Identification Number.  20070=S-COSMOS (Sanders et al.\ 2007); 
61043=SEDS (Ashby et al.\ 2013a); 80057=S-CANDELS (Ashby et al.\ 2015); 90042 \& 10159=SPLASH 
(Steinhardt et al.\ 2014); 11016=SMUVS.}
\tablenotetext{b}{A fourth epoch of PID90042 consisted of just 16 AORs and although it was included in the SMUVS mosaics, it was not separately coadded.}
\end{deluxetable}

To make full use of the unprecedented depth and sensitivity of UltraVISTA's near-IR imaging for 
studies of high-redshift galaxies, deep photometry at longer wavelengths is needed.   
{\sl Spitzer}/IRAC is the obvious facility to provide it.  Indeed, as described below and illustrated by 
Table~\ref{tab:obslog}, different portions of COSMOS have been 
observed with IRAC several times over the course of the {\sl Spitzer} mission.  
The character of these IRAC surveys has varied considerably, 
and includes both wide-and-shallow and narrow-and-deep designs.  Since the 
first visit with IRAC in Cycle 2, the {\sl Spitzer} mission has spent 
nearly 4000\,hr surveying COSMOS, much more than for
 any other IRAC survey completed to date.\footnote{In Cycles 13 and 14,
{\sl Spitzer} began carrying out a final additional deep survey within COSMOS (PID 13094, PI Labb\'e; 1500\,hr) to deepen the coverage between the SMUVS stripes, and (PID 14045, PI Stefanon; 500\,hr), 
to extend the deep coverage to the east and west of the SMUVS stripes, 
creating a single wide-and-deep survey field.  
These observations will be described in future contributions.}   
Roughly 1770\,hr of {\sl Spitzer} time 
were devoted to the new SMUVS observations described here.


\subsection{IRAC Surveys of COSMOS Spanning the Last Decade}
\label{ssec:epochs}

The first IRAC
coverage was obtained during the cryogenic phase of the mission by {\sl Spitzer}-COSMOS 
(S-COSMOS; Sanders et al.\ 2007), which imaged
essentially all of COSMOS with 20\,min total exposure times in all four then-operating IRAC bands.  
Subsequently, relatively small areas within UltraVISTA stripe 2
were imaged during Cycles 6 and 8 of {\sl Spitzer}'s warm mission by the Spitzer Extended Deep
Survey (SEDS; PI Fazio; Ashby et al.\ 2013a) and the {\sl Spitzer}-Cosmic Assembly Near-Infrared
Deep Extragalactic Survey (S-CANDELS; PI Fazio; Ashby et al.\ 2015).   Then in Cycles 9 and 10, the
{\sl Spitzer} Large Area Survey with Hyper-Suprime-Cam (SPLASH; PI Capak; Steinhardt et al.\ 2014)
 imaged almost all of COSMOS much more deeply than S-COSMOS.
The resulting combined
deep coextensive IRAC and UltraVISTA imaging, with photometry spanning many wavebands 
(e.g., Ilbert et al.\ 2010; Laigle et al.\ 2016) proved very useful for identifying high-redshift galaxies
(e.g., Steinhardt et al.\ 2016).  However, even with SPLASH, SEDS, and 
S-COSMOS, the most distant galaxies remained out of reach.  
Thus in Cycle 11, we began a program to cover three of the UltraVISTA
ultra-deep stripes to a much greater and more uniform depth with IRAC, so as to provide 
a much better match to the ground-based near-IR photometry, and over a wide area.  
This program is SMUVS: {\sl Spitzer} Matching survey of the UltraVISTA ultra-deep Stripes, led by 
PI K.\ Caputi.   

\subsection{SMUVS Mapping Strategy}
\label{ssec:strategy}
The UltraVISTA ultra-deep survey covers four parallel stripes of about 0.20\,deg$^2$ each.   
SMUVS covered only stripes 1, 2, and 3, because they benefit from the deepest ancillary data.  
The observing strategy was driven by the need to integrate deeply over these three discontinuous 
fields -- the regions with the deepest $K_s$ imaging -- in as uniform a manner as possible, 
accounting for the different levels of existing coverage.
For example, although S-COSMOS covered all three SMUVS stripes to a uniform depth, Stripes
1 and 3 benefit from fairly deep and uniform coverage by SPLASH.   By design SPLASH did not 
add to the SEDS depths in Stripe 2.  Much of Stripe 2, however, 
was covered to 12\,hr depths by SEDS, a fraction of which was covered with variable but 
long integration times by S-CANDELS, reaching $>100$\,hr in small areas.  The SMUVS observations
were designed to obtain deep coverage over all three stripes by filling in on top of or
adjacent to the existing surveys.

Each SMUVS stripe is roughly 10\arcmin\ wide in Right Ascension, 
and was efficiently mapped with a raster pattern 
having a width equal to two overlapping IRAC fields-of-view.  Given constraints imposed by spacecraft 
scheduling needs, we mapped the stripes in the Declination direction, with small $1\times2$ maps.  
We covered Stripes 1 and 3 respectively with five and six
pairs of such maps (to cover the east and west sides of the stripe).  
Stripe 1 only needed five pointings per half stripe 
because a faulty chip in the VISTA-telescope camera VIRCAM prevented from collecting ultra-deep data 
in the southern part of the stripe.  With its existing deep IRAC coverage, only two
such map pairs were needed to complete Stripe 2.    

Because it lies so close to the ecliptic, each year the COSMOS field is only visible to {\sl Spitzer} 
during two short observing windows roughly 40 days long and 6 months apart, February-March 
and July-August.  SMUVS was designed to
use just the first three available visibility windows, but intense scheduling pressure delayed its completion
until 2017 March.  Thus SMUVS required a total of five visits to COSMOS spread out over more than two
calendar years (Table~\ref{tab:obslog}).  Since the beginning of the mission, the UltraVISTA
ultra-deep stripes therefore have up to 10 distinct imaging epochs in some locations, 
a feature of the dataset that is useful for exploring AGN variability in the near-infrared regime
(S\'anchez et al.\ 2017).

The individual exposures were organized into self-contained segments known as Astronomical
Observing Requests (AORs) roughly six hours long 
to accommodate the downlink schedule.  Each AOR consisted of a sequence of dithered 100\,s 
exposures obtained simultaneously in both operable IRAC detectors.  
All SMUVS AORs used a medium-cycling dithering pattern, which implements half-pixel 
subsampling to cope with cosmic rays, enforce overlap among adjacent 
map positions, and aid in the removal of detector artifacts.  Each map position was observed 
with multiple AORs to accumulate the necessary integration time.  To ensure high redundancy 
the AORs covering any map position were configured with different initial positions for the cycling
dither pattern.  The highly redundant dithering strategy 
also allowed for a thorough sampling of the PSFs.  

\section{Data Reduction}
\label{sec:reduction}

The SMUVS data were reduced using the same procedures that members of our team employed earlier 
with the SEDS and S-CANDELS datasets (Ashby et al.\ 2013a; 2015).  The SMUVS 
reductions differ only in a few minor details.  They are described below.

\subsection{Mosaic Creation}
\label{ssec:coadding}

\begin{figure}[ht!]
\figurenum{2}
\epsscale{2.15}
\plotone{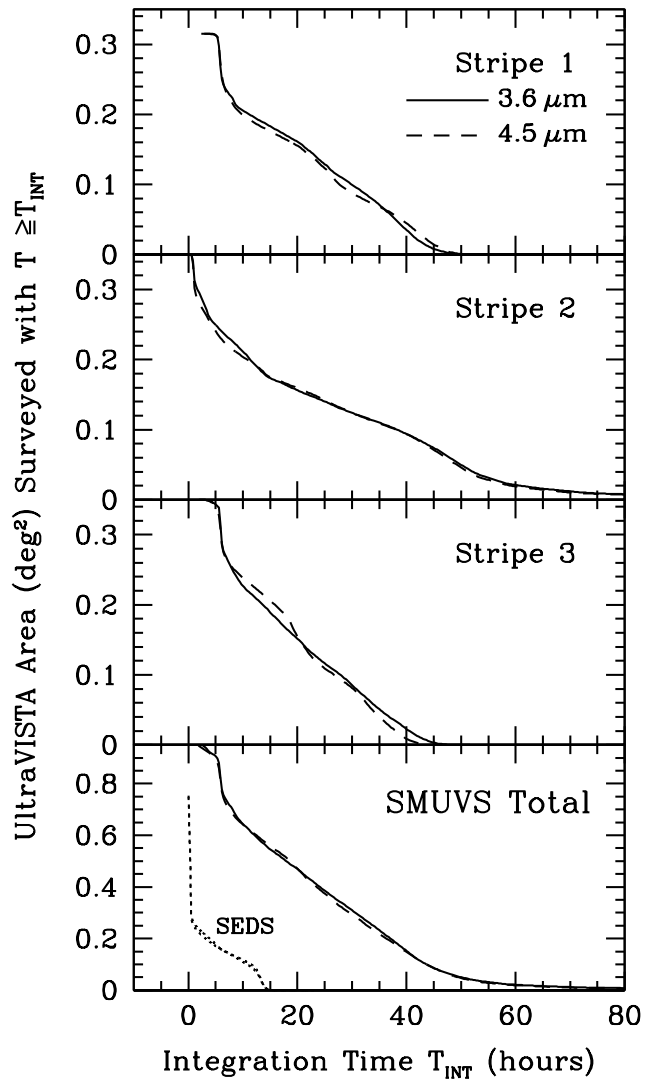}
\caption{Cumulative area coverage as a function of {\sl Spitzer}/IRAC exposure time for SMUVS, including other, earlier observations (Table~\ref{tab:obslog}.)  {\sl Top three panels:} area versus integration time within the three UltraVISTA stripes.  
{\sl Bottom panel:} area versus integration time for all of SMUVS, i.e., the sum of all three stripes.
The data shown are derived from the full-mission IRAC mosaics, 
beginning with S-COSMOS (Sanders et al.\ 2007) during the cryogenic mission 
and continuing through the fifth and final  SMUVS epoch in 2017 April.  At 30\,hr integration times, 
SMUVS covers about 0.3\,deg$^2$, but the coverage is variable.  Nonetheless SMUVS is a significant 
improvement in all respects over, e.g., SEDS, the COSMOS portion of which is shown in the lower panel with dotted lines.
\label{fig:area}}
\end{figure}

After subtracting object-masked median-stacked sky background frames on a per-AOR
basis from all SMUVS exposures to remove long-term residual images, we applied our custom
 column-pulldown corrector to the resulting background-subtracted frames to fix the depressed counts
 in individual array
 columns containing pixels at or near saturation.  We then mosaicked the 
 artifact-corrected exposures, grouped by IRAC band, using {\tt IRACproc} (Schuster et al.\ 2006) 
 within each stripe and epoch.  As was done for SEDS and S-CANDELS, we mosaicked 
 subsets of the exposures to circumvent computer memory limitations and subsequently 
 combined these intermediate-depth mosaics 
 into a single mosaic covering each stripe.  All six SMUVS mosaics were pixellated to 
 0\farcs6 to afford slightly higher effective spatial resolution than the $\sim$1\farcs2 IRAC native pixel size, 
 and were aligned to the tangent-plane projection used by the UltraVISTA collaboration 
 (including Stripes 1 and 3, which do not contain the tangent point).   
 All coextensive non-SMUVS exposures available in the {\sl Spitzer} archive were incorporated into our mosaics
 after processing in the same AOR-based manner as the SMUVS data themselves.
  Thus our final mosaics are full-mission coadds of all IRAC exposures within each 
  UltraVISTA stripe, including data from both the
 cryogenic and warm-mission phases.  The resulting coverage as a function of total integration time
 is shown in Fig.~\ref{fig:area}.

\begin{figure*}
\figurenum{3}
\epsscale{1.2}
\plotone{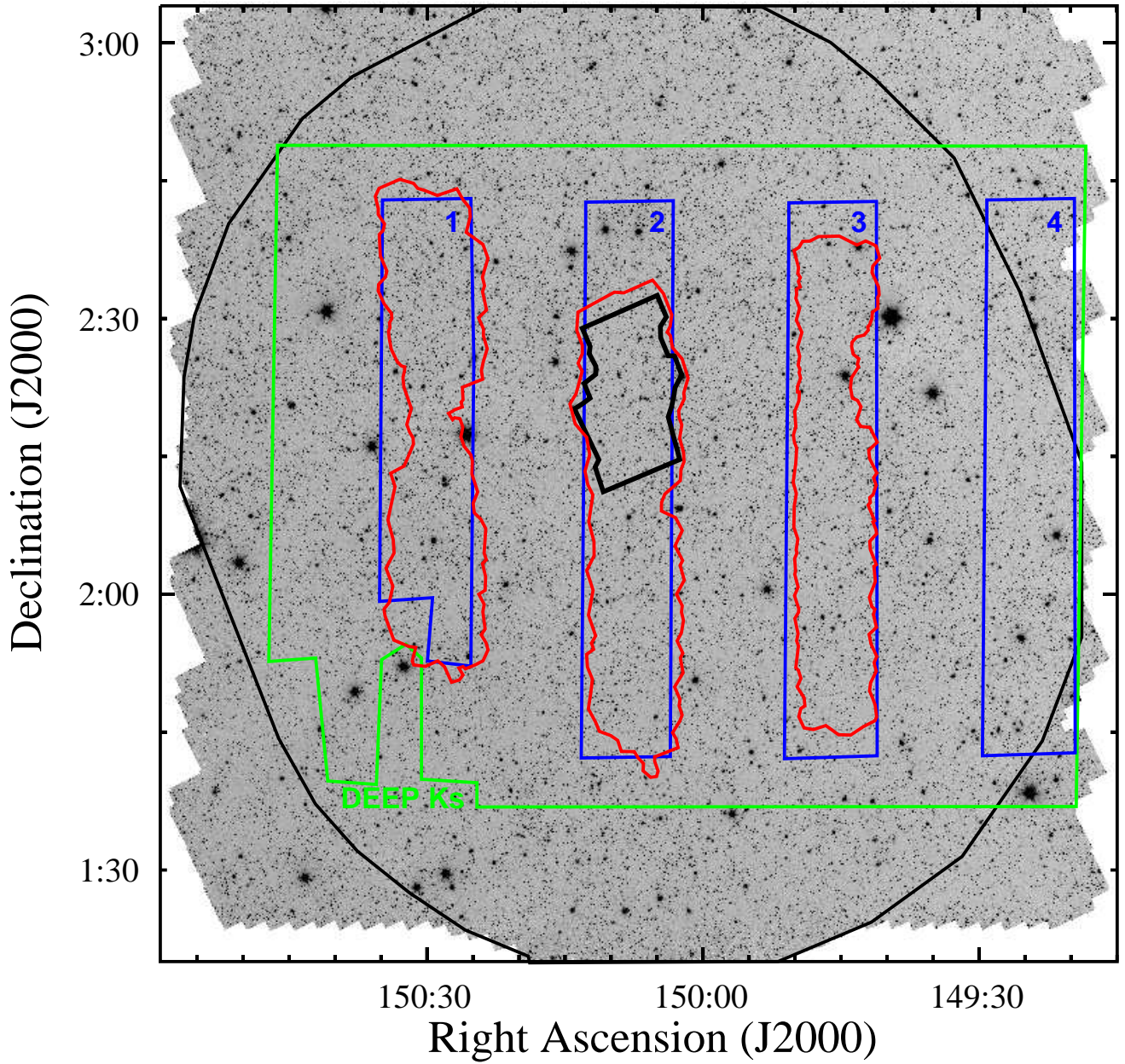}
\caption{Layout of 3.6\,$\mu$m survey coverage within the COSMOS field.  
A higher-resolution version is available in the published version of this article.  The grayscale image 
is a shallow IRAC 3.6\,$\mu$m mosaic built with exposures from all projects listed in 
Table~\ref{tab:obslog} for illustration purposes.  
The linear stretch runs from $-0.05$ to 0.05\,MJy\,sr$^{-1}$.   The entire field was imaged
by S-COSMOS.  The large black ellipse indicates the approximate outer boundary of 
SPLASH-COSMOS.  The green polygon encloses the UltraVISTA deep $K_s$ coverage.  
The four numbered blue polygons are the ultra-deep survey stripes.  
Red contours outline the regions with at least 25\,hr of integration  from SMUVS.  
Outside the red contours the depth of coverage smoothly declines to the $\sim5$\,hr SPLASH-COSMOS
integration times.
The black outline in the center of the field encloses the deep coverage from 
S-CANDELS, i.e., the area with at least 50\,hr integration time.\label{fig:layout1}}
\end{figure*}

\begin{figure*}
\figurenum{4}
\epsscale{1.2}
\plotone{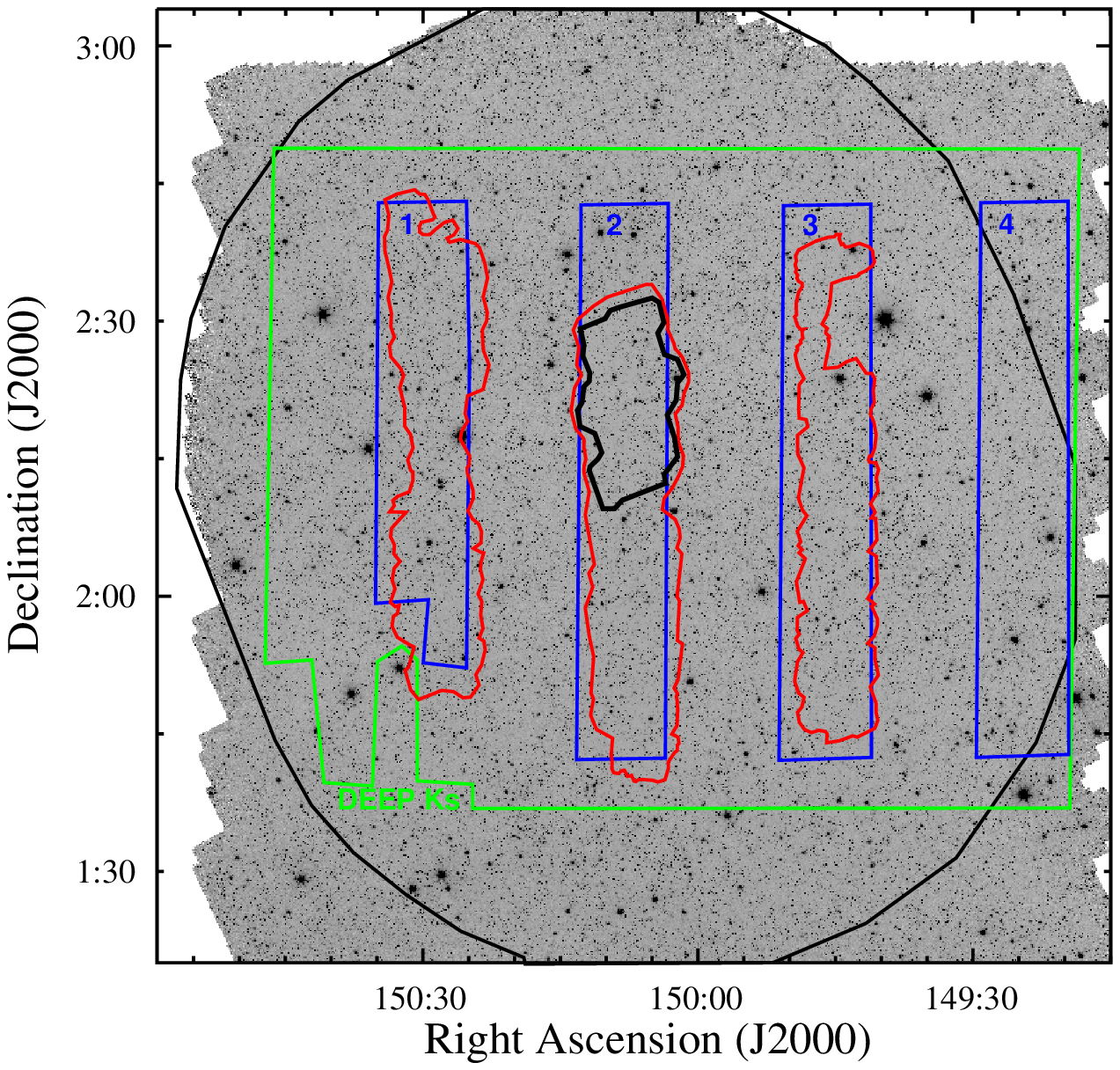}
\caption{As Fig.~\ref{fig:layout1}, but showing the 4.5\,$\mu$m survey coverage.\label{fig:layout2}}
\end{figure*}

 Figures~\ref{fig:layout1} and \ref{fig:layout2} show where the SMUVS coverage is located
 within the COSMOS field.   The SMUVS mosaics and the associated
 coverage maps are all available from the Spitzer Exploration Science Programs 
 website.\footnote{{http://irsa.ipac.caltech.edu/data/SPITZER/docs/spitzer mission/observingprograms/es/}}

\section{SMUVS Catalog Construction}

\subsection{Model PSF Generation}
\label{ssec:PSF}

The SMUVS catalogs (Sec.\ \ref{sec:format}) are based on point spread function (PSF)-fitting techniques 
implemented with {\tt StarFinder} (Diolaiti et al.\ 2000), following analogous procedures 
to those used for SEDS (Ashby et al.\ 2013a).  {\tt StarFinder}
uses scaled model PSFs to estimate source fluxes.   For this reason, the first step
in creating SMUVS catalogs is generating suitable model PSFs.  Unlike S-CANDELS,
which covered too small an area for reliable model PSFs to be constructed, each of
the three SMUVS stripes included many stars suitable for PSF modeling.  
We chose to take advantage of SMUVS' greater area coverage and generate new model PSFs
to optimize our PSF-fitted photometry.

\begin{figure*}
\figurenum{5}
\epsscale{1.2}
\plotone{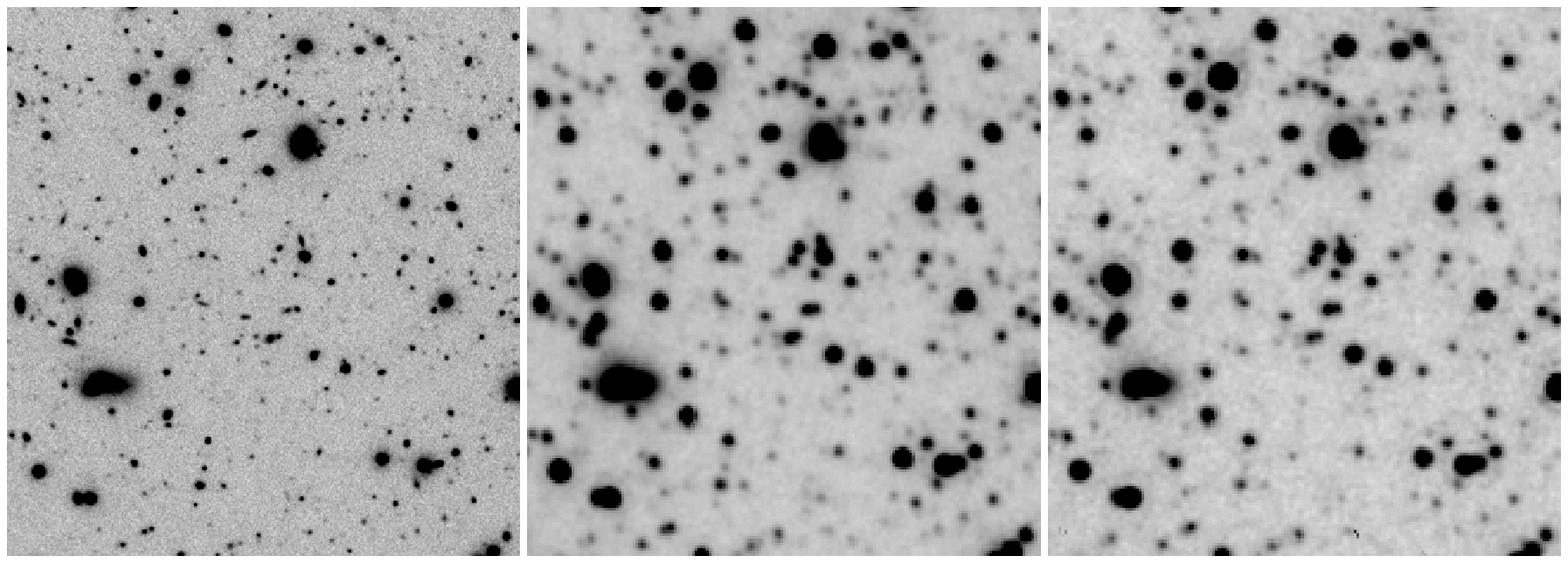}
\caption{Three views of a small but typical SMUVS field to demonstrate the comparable
sensitivities achieved by SMUVS and UltraVISTA.  From left to right the panels show 
mosaics generated with the combined $HK_s$ images from UltraVISTA  (Caputi et al.\ 2017), 
and at 3.6 and 4.5\,$\mu$m for SMUVS.  North is up and east is to the left.  The field shown 
is approximately $100\times120$\,arcsec$^2$, and is located at 10:01:50, +2:00 in stripe 1.  
\label{fig:comparison}}
\end{figure*}

The {\tt StarFinder} algorithm for generating model PSFs can be distilled down to its
essence in three parts.  These are, first, identifying 
isolated, unsaturated field stars, second, generating cutout images centered on those stars and 
cleaning them of nearby contaminating sources, and third, scaling and median-stacking
the cutout images.  All cutouts were centered on the brightest PSF pixel.  
This procedure generated high-dynamic-range model PSFs with relatively high S/N ratios, that
by construction reflect the spacecraft rotation angles at which the individual exposures
were obtained.  There is a limit to the fidelity of this procedure.  Because the SMUVS AORs
consisted of many small, deep maps (i.e., they did not individually 
cover the UltraVISTA stripes uniformly), the ensemble of rotation angles is a function of location within
the stripes.  Our
technique, outlined below, generated 'stripe-average' PSFs that do not fully reflect the small-scale
variations.  The limited visibility of the field, however -- COSMOS is accessible to {\sl Spitzer} only during
windows $\sim$40 days long -- means that the spacecraft can rotate only through a limited
range of position angles, so our approach is a reasonable 
compromise between fidelity and convenience, as we show below.

We used only bright, unsaturated PSF stars observed with at least 25\,hours total
integration time to ensure their images reflected a representative distribution of position angles.  
In stripe 2, with its greater average integration time, we
refined the model PSFs by iterating the {\tt StarFinder} PSF stacking procedure on a
version of the science mosaic from which contaminating field sources had been fitted and subtracted 
on a first pass, down to 5$\sigma$ significance.  The procedure did not noticeably improve
the PSFs for stripes 1 and 3 (it generated faint but broad artifacts in the extreme PSF wings), 
so in these fields the first-pass PSFs were adopted for the final catalogs.  

All PSFs used here were post-processed to improve their suitability for photometry.   
{\tt Starfinder}'s halo-smoothing feature was applied to suppress noise in the PSF outskirts.
In addition, all PSF pixels farther than 64 pixels from the centroid were set to zero, and 
low-level artifacts remaining from the PSF construction process were eliminated by hand.  
These two steps prevented the iterative scaling-and-fitting procedure from introducing spurious
features near the brighter sources.  All PSFs images were subsequently normalized to unity total counts.  
As a sanity check they were then compared visually to the four-epoch PSF images generated 
by Caputi et al.\ (2017) and found to be broadly consistent, but with higher dynamic
ranges.  They were also larger, with 128-pixel diameters  ($76\farcs8$).

\begin{figure*}[ht!]
\figurenum{6}
\epsscale{1.25}
\plotone{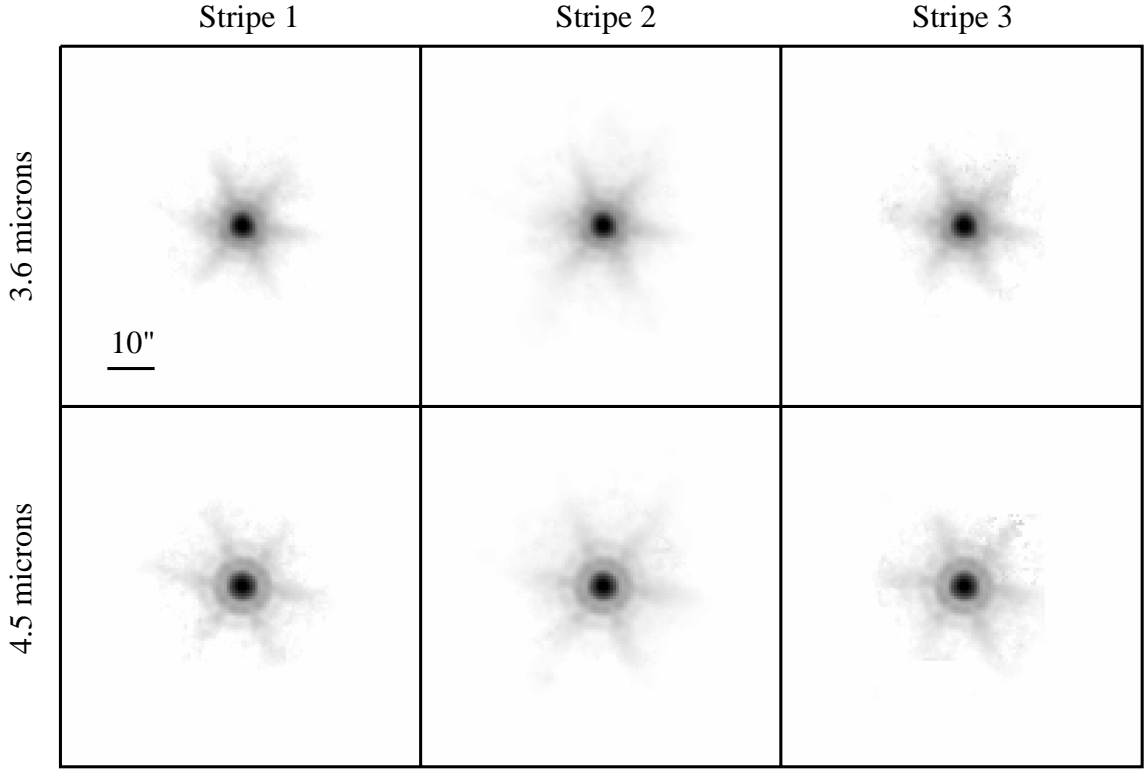}
\caption{Montage of the PSFs derived from the full-mission SMUVS mosaics.
These PSFs were used to construct the SMUVS catalogs.  
From left to right, PSFs are shown for UltraVISTA stripes 1 to 3.  
Each PSF image is 128 0\farcs6 pixels wide.
The upper row shows the 3.6\,$\mu$m PSFs derived by median-stacking of isolated
unsaturated stars all of which were observed for at least 25\,hr.  The lower row shows their
4.5\,$\mu$m equivalents.  The inverse logarithmic stretch ranges from 0 to 0.03 in order to show faint structure
such as the Airy rings; PSF peak values are roughly 0.06 for these normalized PSFs.
\label{fig:PSFs}}
\end{figure*}

Ultimately we generated a total of six model PSF images,
one for every combination of IRAC band and SMUVS stripe.
The SMUVS PSFs are shown in Fig.~\ref{fig:PSFs}.   They have FWHMs of 
approximately 2\arcsec. 

\subsection{Source Extraction}
\label{ssec:catalogs}

To the greatest extent possible the SMUVS photometry was computed in the same
way as was done earlier for SEDS and S-CANDELS, following the standard {\tt StarFinder}
procedure.  In this scheme, the brightest source in the mosaic is identified and fitted with an
appropriately scaled PSF to estimate its brightness.  That source is then subtracted from the
original image, and the process is repeated with the brightest source in the resulting residual 
image.  By looping through this single-source fitting procedure until no significant detections
remain in the residual, {\tt StarFinder} generates a catalog of PSF-fitted estimated fluxes in
brightness order.  We ran iterated this process three times.  On the first pass we set a 
5$\sigma$ detection threshold.  For the second and third passes {\tt StarFinder} estimated the 
RMS from the source-subtracted mosaic, and could reach sources not detected in the
original mosaic.  We also set the detection threshold to 3$\sigma$ for the second and third 
passes to increase the sensitivity.  Regions within $0.7\times$ the PSF FWHM were excluded
from subsequent fits.  Backgrounds were estimated locally for each source,
within square regions $72\times$ the PSF FWHM on a side.  

In all respects our 
{\tt StarFinder} parameters 
were identical to those used for SEDS and S-CANDELS, with one exception.  For 
those earlier efforts, regions nearer than $0.5\times$ the PSF FWHM of detected sources
were excluded from fitting.
Thus some sources in the SEDS and S-CANDELS catalogs may not, if heavily blended with a 
brighter companion, appear in the SMUVS catalogs, which are slightly more resistant to shredding of bright sources.

After the iterated fitting procedure was carried out on all three SMUVS stripes within the blue boundaries
indicated in Figs.~\ref{fig:layout1} and \ref{fig:layout2}, aperture photometry was acquired at the 
positions of all {\tt StarFinder}-detected sources.  This was done by adding the scaled PSFs back
into the residual images and photometering the resulting 'reconstituted' sources within apertures of diameters
2\farcs4, 3\farcs6, 4\farcs8, 6\farcs0, 7\farcs2, and 12\farcs0.  This technique permitted the photometry
to be measured with less contamination from nearby sources and, because the nearby {\tt StarFinder}-detected
sources were by construction removed from the residual image used.  The {\tt StarFinder} PSFs were used 
to estimate and correct for flux falling outside the apertures.    

The two resulting single-band IRAC catalogs for each stripe were then combined into a single two-band catalog 
using a position match 
with a 1\arcsec\ search radius.  The 1\arcsec\ radius was selected because it is smaller than the FWHM 
of the IRAC PSF in either band, but larger than the 0\farcs7 exclusion radius around detected sources. 
After inspecting the catalogs, we chose to retain significant unmatched 
sources in order to improve the catalog completeness.  Thus at faint levels, a significant fraction of the
SMUVS detections are formally detected in only one IRAC band.
 
 In summary, we have constructed one two-band position-matched catalog for each of the three SMUVS stripes.
The catalogs are presented in Tables~\ref{tab:stripe1}, \ref{tab:stripe2}, and \ref{tab:stripe3}.  The catalogs
contain a total of about 356,000 sources down to 4$\sigma$ limits of roughly 25.0\,AB mag in
both IRAC bands.  The sensitivity limit accounts for both instrumental effects and the 
effects of source confusion.

 \subsection{The Impact of Source Confusion}
 
 Source confusion is significant at SMUVS depths.  Indeed, it had a measurable (if marginal) 
 impact even on the shallow counts in the SSDF 
 (Ashby et al.\ 2013b).   SMUVS is significantly deeper than SEDS (designed integration time 
 of 12\,hr, Ashby et al.\ 2013a; Fig.~\ref{fig:area}, bottom panel).  
  We identified about 350,000 significant sources in the 0.66\,deg$^2$ 
 total SMUVS area,  equivalent to roughly 10 beams per source, a level well above the 
 40 beams per source criterion for the onset of source confusion given in Rowan-Robinson (2001).
 Including fainter (less significant but nonetheless real) objects, of course, would raise the estimated  
 source confusion accordingly.  Accounting for source confusion was a primary 
 motive behind our decision to use 
 {\tt StarFinder} in this work. 
 
 We used the COSMOS S-CANDELS observations to estimate the impacts of
 source confusion on SMUVS.  The S-CANDELS survey provides a reliable means of dealing 
 quantitatively with SMUVS source confusion for four reasons.  First, S-CANDELS reaches fainter 
 flux levels than SMUVS along the same general line of sight, and therefore accurately accounts 
 for the behavior of real sources that
 SMUVS cannot detect reliably.  Second, {\sl Spitzer's} short visibility windows for COSMOS mean that
 all IRAC observations of the field were taken at nearly identical spacecraft rotation angles, so the
 resulting PSFs are likewise nearly identical.  Third, the SMUVS and S-CANDELS mosaics' 
 pixellation and tangent points were identical by construction.  Thus the  {\tt StarFinder} source
 extraction simulations performed on the S-CANDELS mosaics by Ashby et al.\ (2015) 
 are representative of the SMUVS source extractions at the same flux levels.
 
Source confusion dominates the photometric uncertainties for faint IRAC sources.  
Thus, the total uncertainties do not integrate down as
the square root of the integration time as they would in the absence of confusion noise.
 As shown in Ashby et al.\ (2015), Fig.\ 15 and Table 3, the total uncertainty for 
COSMOS 3.6\,$\mu$m sources photometered with {\tt StarFinder} is 0.1\,mag at 21.25\,AB mag, 
but this only 0.02\,mag greater (i.e., only roughly 20\% more uncertain) than for 
 sources that are brighter by a full magnitude or even more.  
 
The tendency for  confusion-dominated IRAC photometric uncertainties to grow slowly
toward faint magnitudes has a somewhat non-intuitive consequence for source significance.
For SMUVS in particular, a 25\,mag (0.13\,$\mu$Jy) source is a 4$\sigma$ detection.  But 
a source half as bright (25.75\,mag, 0.065\,$\mu$Jy) is not a 2$\sigma$ detection -- it is closer
to 3$\sigma$.  Users should bear this behavior in mind when using the SMUVS catalogs.

\begin{deluxetable*}{cccccccccccrc}
\tabletypesize{\scriptsize}
\tablecaption{Full-Depth Source Catalog for SMUVS Stripe 1\label{tab:stripe1}}
\tablehead{
\colhead{Object} &
\colhead{RA,Dec} &
\colhead{3.6\,$\mu$m AB Magnitudes\tablenotemark{a}} &
\colhead{3.6\,$\mu$m Unc.\tablenotemark{b}} &
\colhead{3.6\,$\mu$m Coverage\tablenotemark{c}} &
\colhead{3.6\,$\mu$m Flag\tablenotemark{d}} \\ 
&
\colhead{(J2000)}&
\colhead{4.5\,$\mu$m AB Magnitudes} &
\colhead{4.5\,$\mu$m Unc.} &
\colhead{4.5\,$\mu$m Coverage\tablenotemark{e}} &
\colhead{4.5\,$\mu$m Flag\tablenotemark{f}} \\
}
\startdata
SMUVS J100143.20+021729.0 & 150.43001,2.29139 &  \ 9.64  \ 9.60  \ \ 9.52 \ 9.48 \ \ 9.45  \ 9.44  \ \ 9.44 & 0.03 &  748 & 1 \\
 & & 10.13 10.08  \ 9.96  \ 9.92 \ \ 9.91  \ 9.90  \ \ 9.89 & 0.03 &  885 & 1 \\
SMUVS J100210.51+015212.0 & 150.54377,1.87000 & 10.53 10.49 10.38 10.33 10.30 10.29 10.28 & 0.03 &  618 & 1 \\
 & & 11.05 11.00 10.89 10.85 10.83 10.82 10.81 & 0.03 &  622 & 1 \\
SMUVS J100223.99+021604.6 & 150.59996,2.26795 & 10.54 10.51 10.40 10.36 10.35 10.34 10.33 & 0.03 &  201 & 1 \\
 & & 11.07 11.03 10.92 10.88 10.86 10.85 10.85 & 0.03 &  179 & 1 \\
SMUVS J100157.37+020556.3 & 150.48904,2.09898 & 11.93 11.89 11.80 11.77 11.75 11.75 11.75 & 0.03 & 1362 & 0 \\
 & & 12.56 12.52 12.43 12.39 12.37 12.36 12.36 & 0.03 & 1420 & 0 \\
SMUVS J100142.19+015320.0 & 150.42579,1.88888 & 12.40 12.36 12.29 12.25 12.23 12.23 12.23 & 0.03 &  884 & 0 \\
 & & 13.11 13.07 12.95 12.89 12.87 12.86 12.86 & 0.03 & 1054 & 0 \\
SMUVS J100130.37+023616.1 & 150.37654,2.60446 & 12.41 12.37 12.29 12.26 12.24 12.24 12.24 & 0.03 &  522 & 0 \\
 & & 13.09 13.05 12.93 12.88 12.85 12.84 12.84 & 0.03 &  695 & 0 \\
SMUVS J100152.83+021233.5 & 150.47012,2.20932 & 12.60 12.57 12.48 12.45 12.43 12.42 12.42 & 0.03 & 1571 & 0 \\
 & & 13.22 13.18 13.07 13.03 13.01 13.00 13.00 & 0.03 & 1590 & 0 \\
SMUVS J100214.05+022416.0 & 150.55853,2.40444 & 12.64 12.61 12.54 12.50 12.49 12.48 12.48 & 0.03 &  402 & 0 \\
 & & 13.31 13.27 13.16 13.10 13.07 13.06 13.05 & 0.03 &  363 & 0 \\
SMUVS J100125.93+020109.5 & 150.35805,2.01930 & 12.68 12.65 12.57 12.55 12.53 12.53 12.52 & 0.03 &  268 & 0 \\
 & & 13.44 13.39 13.27 13.20 13.17 13.16 13.15 & 0.03 &  301 & 0 \\
\enddata
\tablecomments{The SMUVS catalog of IRAC-detected sources in Stripe 1.
The sources are listed in magnitude order.
This table is available in its entirety in a machine-readable form in the online
journal.  A portion is shown here for guidance regarding its form and content.
}
\tablenotetext{a}{The PSF-fitted magnitude is given first, and the magnitudes
given after are measured in apertures of 2\farcs4, 3\farcs6, 4\farcs8, 6\farcs0,
7\farcs2, and 12\farcs0 diameter, corrected to total.}
\tablenotetext{b}{Uncertainties given are 1$\sigma$, expressed in magnitudes, and apply to the 2\farcs4
diameter aperture magnitudes.}
\tablenotetext{c}{Depth of coverage expressed in units of 100\,s IRAC 3.6\,$\mu$m frames that observed the source.}
\tablenotetext{d}{Flag indicating possible corrupted 3.6\,$\mu$m photometry (if equal to 1) due to proximity to a bright star, or (if equal to 2) a single-band detection.}
\tablenotetext{e}{Depth of coverage expressed in units of 100\,s IRAC 4.5\,$\mu$m frames that observed the source.}
\tablenotetext{f}{Flag indicating possible corrupted 4.5\,$\mu$m photometry (if equal to 1) due to proximity to a bright star, or (if equal to 2) a single-band detection.}
\end{deluxetable*}

\begin{deluxetable*}{cccccccccccrc}
\tabletypesize{\scriptsize}
\tablecaption{Full-Depth Source Catalog for SMUVS Stripe 2\label{tab:stripe2}}
\tablehead{
\colhead{Object} &
\colhead{RA,Dec} &
\colhead{3.6\,$\mu$m AB Magnitudes\tablenotemark{a}} &
\colhead{3.6\,$\mu$m Unc.\tablenotemark{b}} &
\colhead{3.6\,$\mu$m Coverage\tablenotemark{c}} &
\colhead{3.6\,$\mu$m Flag\tablenotemark{d}} \\ 
&
\colhead{(J2000)}&
\colhead{4.5\,$\mu$m AB Magnitudes} &
\colhead{4.5\,$\mu$m Unc.} &
\colhead{4.5\,$\mu$m Coverage\tablenotemark{e}} &
\colhead{4.5\,$\mu$m Flag\tablenotemark{f}} \\
}
\startdata
SMUVS J100009.66+022349.0 & 150.04023,2.39693 & 11.11 11.08 11.02 10.98 10.97 10.97 10.97 & 0.03 & 1335 & 1 \\
 & & 11.71 11.66 11.55 11.52 11.51 11.50 11.49 & 0.03 &  496 & 1 \\
SMUVS J100003.58+015044.9 & 150.01493,1.84579 & 11.60 11.56 11.49 11.46 11.44 11.43 11.42 & 0.03 &  436 & 1 \\
 & & 12.15 12.09 11.97 11.91 11.88 11.85 11.83 & 0.03 &  556 & 1 \\
SMUVS J100057.12+023719.3 & 150.23798,2.62204 & 11.69 11.64 11.63 11.62 11.59 11.56 11.49 & 0.03 &  185 & 1 \\
 & & 11.52 11.47 11.36 11.31 11.29 11.28 11.27 & 0.03 &   58 & 1 \\
SMUVS J100042.71+023941.6 & 150.17796,2.66155 & 11.88 11.83 11.72 11.70 11.69 11.67 11.62 & 0.03 &  375 & 1 \\
 & & 11.90 11.85 11.75 11.72 11.70 11.69 11.68 & 0.03 &  288 & 1 \\
SMUVS J100028.36+023926.1 & 150.11818,2.65725 & 12.10 12.05 11.96 11.92 11.90 11.88 11.85 & 0.03 &  489 & 1 \\
 & & 12.25 12.20 12.14 12.11 12.10 12.09 12.08 & 0.03 &  274 & 1 \\
SMUVS J100002.36+023259.5 & 150.00982,2.54987 & 12.65 12.62 12.55 12.53 12.52 12.51 12.52 & 0.03 &   36 & 1 \\
 & & 13.50 13.44 13.35 13.30 13.28 13.27 13.26 & 0.03 &   41 & 1 \\
SMUVS J100032.55+020825.8 & 150.13564,2.14049 & 12.70 12.67 12.61 12.58 12.56 12.55 12.55 & 0.03 & 1535 & 1 \\
 & & 13.21 13.17 13.08 13.04 13.01 13.00 13.00 & 0.03 & 1345 & 1 \\
SMUVS J100024.41+024422.6 & 150.10170,2.73961 & 12.92 12.87 12.79 12.74 12.71 12.69 12.67 & 0.03 &  292 & 1 \\
 & & 13.03 12.98 12.88 12.85 12.83 12.82 12.81 & 0.03 &  179 & 1 \\
SMUVS J100057.88+023535.6 & 150.24118,2.59322 & 13.50 13.38 13.18 13.06 12.99 12.96 12.91 & 0.03 &  127 & 1 \\
 & & 13.07 13.03 12.94 12.89 12.86 12.84 12.83 & 0.03 &   32 & 1 \\
\enddata
\tablecomments{The SMUVS catalog of IRAC-detected sources in Stripe 1.
The sources are listed in magnitude order.
This table is available in its entirety in a machine-readable form in the online
journal.  A portion is shown here for guidance regarding its form and content.
}
\tablenotetext{a}{The PSF-fitted magnitude is given first, and the magnitudes
given after are measured in apertures of 2\farcs4, 3\farcs6, 4\farcs8, 6\farcs0,
7\farcs2, and 12\farcs0 diameter, corrected to total.}
\tablenotetext{b}{Uncertainties are 1$\sigma$, expressed in magnitudes, and apply to the 2\farcs4
diameter aperture magnitudes.}
\tablenotetext{c}{Depth of coverage expressed in units of 100\,s IRAC 3.6\,$\mu$m frames that observed the source.}
\tablenotetext{d}{Flag indicating possible corrupted 3.6\,$\mu$m photometry (if equal to 1) due to proximity to a bright star, or (if equal to 2) a single-band detection.}
\tablenotetext{e}{Depth of coverage expressed in units of 100\,s IRAC 4.5\,$\mu$m frames that observed the source.}
\tablenotetext{f}{Flag indicating possible corrupted 4.5\,$\mu$m photometry (if equal to 1) due to proximity to a bright star, or (if equal to 2) a single-band detection.}
\end{deluxetable*}

\begin{deluxetable*}{cccccccccccrc}
\tabletypesize{\scriptsize}
\tablecaption{Full-Depth Source Catalog for SMUVS Stripe 3\label{tab:stripe3}}
\tablehead{
\colhead{Object} &
\colhead{RA,Dec} &
\colhead{3.6\,$\mu$m AB Magnitudes\tablenotemark{a}} &
\colhead{3.6\,$\mu$m Unc.\tablenotemark{b}} &
\colhead{3.6\,$\mu$m Coverage\tablenotemark{c}} &
\colhead{3.6\,$\mu$m Flag\tablenotemark{d}} \\ 
&
\colhead{(J2000)}&
\colhead{4.5\,$\mu$m AB Magnitudes} &
\colhead{4.5\,$\mu$m Unc.} &
\colhead{4.5\,$\mu$m Coverage\tablenotemark{e}} &
\colhead{4.5\,$\mu$m Flag\tablenotemark{f}} \\
}
\startdata
SMUVS J095838.29+023010.2 & 149.65956,2.50284 &  \ 8.86  \ 8.82  \ 8.76  \ 8.72  \ 8.71  \ 8.70  \ 8.69 & 0.03 &  539 & 1 \\
 & &  \ 9.43  \ 9.38  \ 9.29  \ 9.25  \ 9.23 \  9.22  \ 9.21 & 0.03 &  596 & 1 \\
SMUVS J095858.28+022346.9 & 149.74282,2.39637 & 10.72 10.68 10.59 10.54 10.52 10.51 10.50 & 0.03 & 1179 & 1 \\
 & & 11.20 11.14 11.03 10.98 10.96 10.94 10.93 & 0.03 & 1041 & 1 \\
SMUVS J095932.36+020032.7 & 149.88482,2.00909 & 11.08 11.05 10.96 10.92 10.91 10.90 10.89 & 0.03 &  256 & 1 \\
 & & 11.48 11.44 11.36 11.33 11.31 11.30 11.29 & 0.03 &  161 & 1 \\
SMUVS J095852.55+023748.1 & 149.71896,2.63002 & 11.28 11.24 11.16 11.13 11.11 11.11 11.11 & 0.03 &  988 & 1 \\
 & & 11.82 11.77 11.68 11.64 11.61 11.60 11.60 & 0.03 &  956 & 1 \\
SMUVS J095839.21+020905.6 & 149.66337,2.15154 & 12.60 12.56 12.50 12.47 12.45 12.45 12.44 & 0.03 &  411 & 0 \\
 & & 13.30 13.25 13.12 13.06 13.04 13.02 13.01 & 0.03 &  491 & 0 \\
SMUVS J095833.72+014348.5 & 149.64051,1.73014 & 12.71 12.69 12.62 12.59 12.57 12.57 12.57 & 0.03 &  215 & 1 \\
 & & 13.53 13.48 13.34 13.28 13.25 13.24 13.23 & 0.03 &  213 & 1 \\
SMUVS J095858.83+013746.1 & 149.74512,1.62947 & 12.89 12.86 12.79 12.76 12.74 12.74 12.74 & 0.03 &  184 & 1 \\
 & & 13.69 13.63 13.51 13.44 13.41 13.39 13.38 & 0.03 &  189 & 1 \\
SMUVS J095920.69+022819.0 & 149.83621,2.47194 & 12.92 12.89 12.82 12.79 12.77 12.77 12.76 & 0.03 &  669 & 0 \\
 & & 13.64 13.59 13.47 13.41 13.38 13.37 13.35 & 0.03 &  475 & 0 \\
SMUVS J095908.29+015732.6 & 149.78455,1.95906 & 12.94 12.91 12.84 12.81 12.79 12.79 12.79 & 0.03 & 1419 & 0 \\
 & & 14.78 14.58 14.32 14.13 13.97 13.90 13.82 & 0.03 &  860 & 0 \\
\enddata
\tablecomments{The SMUVS catalog of IRAC-detected sources in Stripe 1.
The sources are listed in magnitude order.
This table is available in its entirety in a machine-readable form in the online
journal.  A portion is shown here for guidance regarding its form and content.
}
\tablenotetext{a}{The PSF-fitted magnitude is given first, and the magnitudes
given after are measured in apertures of 2\farcs4, 3\farcs6, 4\farcs8, 6\farcs0,
7\farcs2, and 12\farcs0 diameter, corrected to total.}
\tablenotetext{b}{Uncertainties given are 1$\sigma$, expressed in magnitudes, and apply to the 2\farcs4
diameter aperture magnitudes.}
\tablenotetext{c}{Depth of coverage expressed in units of 100\,s IRAC 3.6\,$\mu$m frames that observed the source.}
\tablenotetext{d}{Flag indicating possible corrupted 3.6\,$\mu$m photometry (if equal to 1) due to proximity to a bright star, or (if equal to 2) a single-band detection.}
\tablenotetext{e}{Depth of coverage expressed in units of 100\,s IRAC 4.5\,$\mu$m frames that observed the source.}
\tablenotetext{f}{Flag indicating possible corrupted 4.5\,$\mu$m photometry (if equal to 1) due to proximity to a bright star, or (if equal to 2) a single-band detection.}
\end{deluxetable*}

\section{Catalog Format}
\label{sec:format}

Each SMUVS catalog follows an identical format.  All IRAC sources detected with 4$\sigma$ significance
in at least one IRAC band are included.  The largest catalog section comes first, and lists all sources detected
in both IRAC bands in brightness order.  The entries for sources detected
at 3.6\,$\mu$m but not 4.5\,$\mu$m, and also conversely, appear later.  Invalid measurements are indicated with large
negative numbers throughout.

The entry for each source includes its name and {\tt StarFinder}-derived position.  The positions given are those
measured at 3.6\,$\mu$m except for sources not detected in that band.  For those sources the position measured
at 4.5\,$\mu$m is given instead.

Seven photometric measurements are given in each band for each detection.  The first entry is always the PSF-fitted
magnitude.  The next six measurements are aperture magnitudes as described earlier.  All photometry is stated in
AB terms.  The aperture photometry is corrected to total magnitudes following the same procedure 
used in Ashby et al.\ (2013a) in order to account for imperfect measurement of sky backgrounds in 
such a dense field.  For SMUVS,
the COSMOS-specific corrections were applied (Ashby et al.\ 2015, Fig.\ 15).

Uncertainty estimates are given for the 2\farcs4 diameter aperture photometry.  These estimates are indicative of
the uncertainties obtained for the PSF-fitted magnitudes as well, but should be regarded as  
underestimates for larger diameter apertures.  Wider apertures suffer from two problems.  
First, the wider apertures can encompass extraneous features 
(e.g., faint undetected objects, artifacts, residuals from brighter nearby sources) that will reduce the precision
of the photometry and bias it toward brighter magnitudes.  
Second, wider apertures necessarily have a higher contribution from shot noise.

The numbers of individual IRAC exposures taken at every source position, inferred from the coverage maps
generated during mosaicking, are given in terms of 100\,s exposures.  The numbers given 
are measured at the cataloged positions of the sources.  Because of artifact correction and cosmic
ray rejection, these numbers can vary considerably even on arcsecond scales.  

The data quality flags are described in Sec.~\ref{sec:validation}.

\section{Catalog Validation}
\label{sec:validation}

The astrometric solution for IRAC is tied to the known positions of relatively bright point sources 
in the Two Micron All Sky Survey (2MASS; Skrutskie et al.\ 2006) Point Source Catalog.  
Recently, however, that astrometric solution was improved in two ways (Lowrance et al.\ 2016):  
first, by implementing a fifth-order polynomial to account for optical distortion; second, by accounting for 
the proper motions of bright 2MASS sources (which are significant for 22\% of the 2MASS stars used in
the pointing refinement) using the UCAC4 catalog (Zacharias et al.\ 2013).  For SMUVS, the updated
astrometric solution was used, so we verified the SMUVS astrometry against both  
2MASS and SEDS -- in other words, against the extremes of wide/shallow
and deep/narrow coextensive observations in similar wavebands, and using IRAC astrometry 
measured with the earlier, third-order distortion correction.  The positions of
SMUVS sources are consistent with those from SEDS to within about 0\farcs12.  Relative to 2MASS, the SMUVS source
positions match to within about 0\farcs18 arcsec, consistent with what has been seen in earlier IRAC surveys.

\begin{figure*}[ht!]
\figurenum{7}
\epsscale{1.1}
\plotone{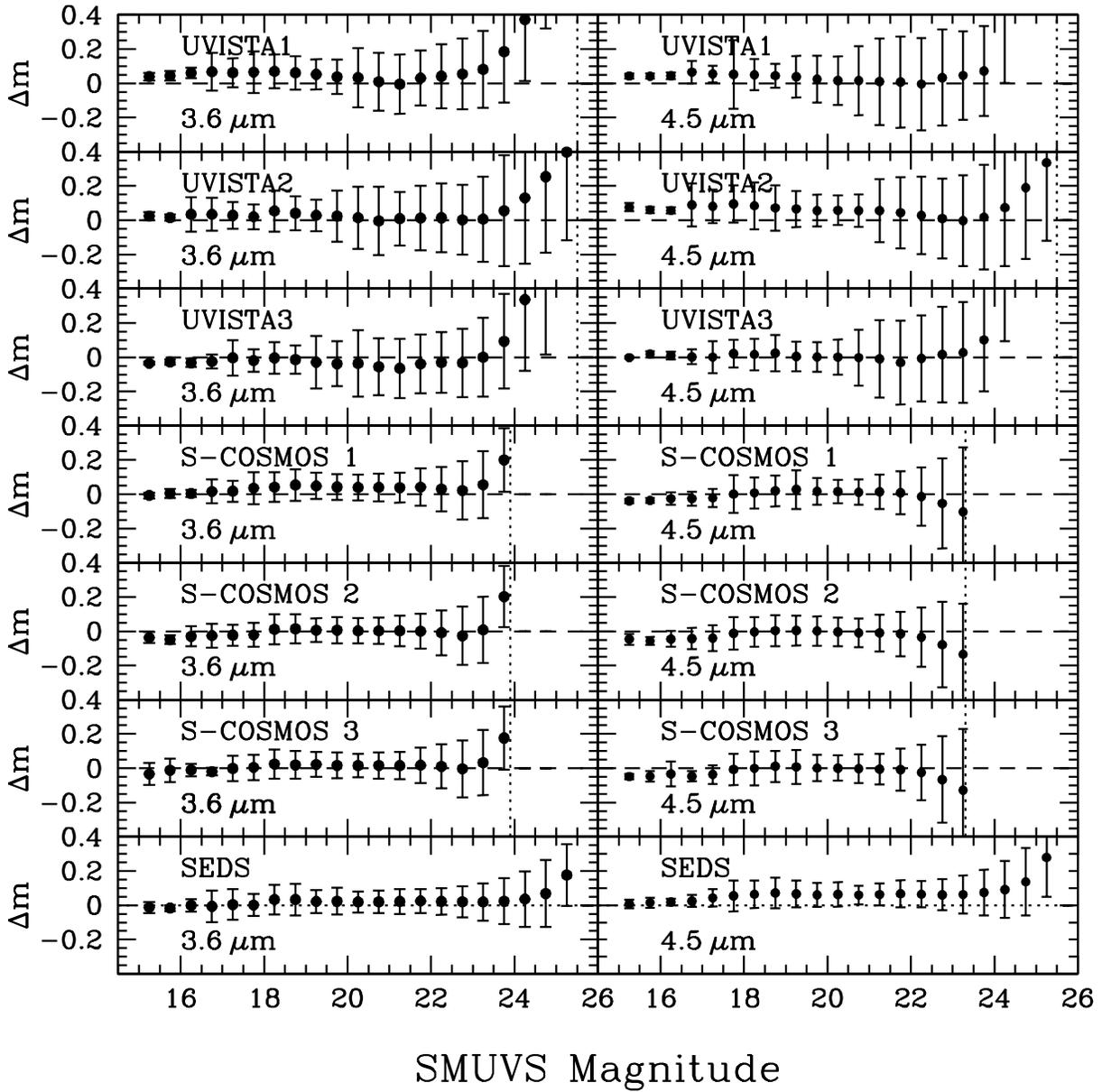}
\caption{Comparisons of SMUVS photometry to previously published results.  All points indicate means of 
magnitude differences for position-matched sources, measured in bins 0.5\,mag wide.  
The error bars are $1\sigma$.  {\sl Top three rows:} SMUVS photometry for stripes 1, 2, and 3 separately compared to that from UltraVISTA using $HK_s$ priors, from Deshmukh et al.\ (2018).  Vertical dashed lines indicate the UltraVISTA 80\% completeness limit.
{\sl Next three rows:} SMUVS photometry compared to coextensive 
S-COSMOS 1\farcs9 diameter aperture photometry (Sanders et al.\  2007) down to the S-COSMOS detection thresholds, indicated by the vertical dashed lines.
{\sl Bottom row:} SMUVS compared to SEDS (Ashby et al.\ 2013a).  The comparison to S-COSMOS was based
on SMUVS 2\farcs4 diameter aperture photometry, and the comparison to SEDS was based on SMUVS and SEDS bias-corrected PSF-fitted magnitudes.
\label{fig:compare}}
\end{figure*}

Fig.~\ref{fig:compare} compares SMUVS photometry to that from SEDS, S-COSMOS, and 
Deshmukh et al.\ (2018).  In all instances, the SMUVS photometry is consistent
with previous measurements within the uncertainties, but the comparison reveals some systematic 
differences among the datasets, which are described here.

{\sl Comparison to Deshmukh et al.\ (2018):} the SMUVS photometry is compared to that of Deshmukh et al.\ (2018) in all three UltraVISTA stripes and both IRAC bands in the top three rows of 
Fig.~\ref{fig:compare}.  Like SMUVS, Deshmukh et al.\ employ a PSF-fitting technique to 
photometer sources in the IRAC bands, but unlike SMUVS their photometry is measured at the
positions of sources detected in a suite of very deep ground-based mosaics built by coadding 
exposures in the $J$ and $K_s$ bands.  SMUVS sources were matched to 
those of Deshmukh et al.\ if their positions were coincident to within 0\farcs4.  All SMUVS sources
 brighter than 25.5\,mag at 3.6 and 4.5\,$\mu$m, were considered.  In addition, we required 
 that the measured IRAC $[3.6]-[4.5]$ color in both catalogs had to agree to within
 0.2\,mag.  This was done to help ensure that the photometry was compared for the same sources, 
 which otherwise would have been problematic because the IRAC sources may resolve into multiple
 objects in the $HKs$-selected catalog.
 
 For bright SMUVS sources (i.e., $[3.6]=[4.5]<16$\,mag) the scatter in the comparison is very small 
 in all three stripes and both bands, and appears dominated by systematic effects that are 
 comparable to the roughly 3\% uncertainty in the absolute calibration of the IRAC.  This behavior
 is apparent in comparisons to SEDS and S-COSMOS as well.  For sources fainter than 16\,mag,
 considerably more scatter is apparent, but the mean deviations from zero difference 
 (SMUVS-Deshmukh) tend to be comparable to the uncertainty in the absolute calibration error
 down to about 24\,mag.  For sources fainter than 24\,mag, the difference between SMUVS and 
 Deshmukh et al.\ is positive and larger than the systematic errors.  We cannot definitively 
 ascribe a cause to the discrepancy, but we speculate that it arises because the two catalogs are
 selected in different wavebands.  Faint SMUVS sources may be systematically absent from 
 Deshmukh et al. (which selects on $HK_s$), but may nonetheless satisfy our simple position and 
color-matching criteria, distorting the comparison in subtle ways.

{\sl Comparison to S-COSMOS:} the SMUVS photometry is compared to that of S-COSMOS in
rows 2-4 of Fig.~\ref{fig:compare}, again for sources matched to within 0\farcs4.  The comparison
was done separately in the IRAC bands, down to the S-COSMOS detection limits of 1.0 and 
1.7\,$\mu$Jy in the 3.6 and 4.5\,$\mu$m bands, respectively (23.9 and 23.3\,AB mag).   
The comparison was based on S-COSMOS 1\farcs9 diameter aperture magnitudes, corrected
to total magnitudes as specified in the S-COSMOS IRAC data delivery README file.  Only 
S-COSMOS objects
with data quality flags set to zero (i.e., good data) were used in the comparison.

The agreement between SMUVS and S-COSMOS for sources brighter than 23.5\,mag is on average
better than the 3\% uncertainty in the absolute IRAC calibration.  S-COSMOS sources fainter than 23.5\,mag at 3.6\,$\mu$m are systematically brighter on average in S-COSMOS than in SMUVS.  
This effect is not seen for the 4.5\,$\mu$m S-COSMOS sources.
 
 {\sl Comparison to SEDS:}  the SMUVS photometry is compared to that of SEDS in the bottom 
 row of Fig.~\ref{fig:compare}.  For this comparison we used the 2\farcs4 diameter aperture 
 magnitudes, corrected to total, from both SEDS and SMUVS.  The agreement between SMUVS 
 and SEDS is excellent at 3.6\,$\mu$m.  At 4.5\,$\mu$m, an offset of about 0.06\,mag is detected, 
 in the sense that the SMUVS photometry is on average systematically 0.06\,mag fainter for
 sources fainter than about 18\,mag.  The origin of this systematic offset is not understood.
 
The bottom row of Fig.~\ref{fig:compare} compares SMUVS sources in a single 0.5\,mag bin fainter 
than our nominal 4$\sigma$ cutoff at 25\,mag.  Such sources appear on average
to be brighter by about 0.2\,mag in SEDS than in SMUVS, with a comparable uncertainty.   
The reason for the difference is not entirely clear, because at this magnitude 
confusion noise dominates the uncertainties in both catalogs.  We speculate that it 
may result from the slightly different selection function used for SMUVS.  
Whereas the SMUVS catalogs include single-band $4\sigma$ detections but discard faint
off-band detections, the SEDS catalogs include (somewhat less significant) detections in both 
bands.  This could lead to a situation where sources  in the shallower SEDS mosaics 
tend to be boosted by noise fluctuations slightly more often than for SMUVS, resulting
in the red average color seen in the faintest bins of the bottom row Fig.~\ref{fig:compare}.

\begin{figure*}[ht!]
\figurenum{8}
\epsscale{1.1}
\plotone{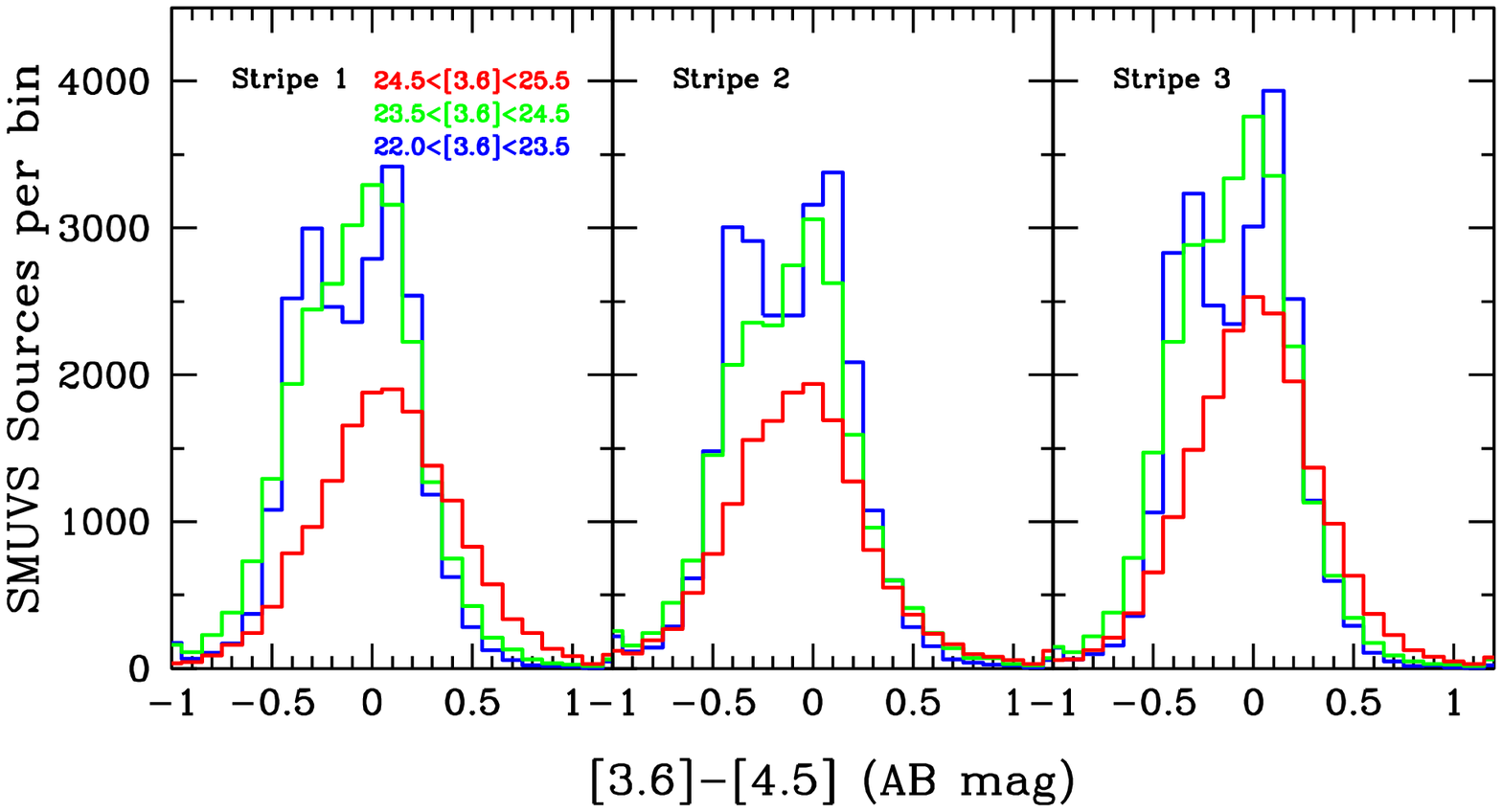}
\caption{The IRAC color distributions of sources detected in both warm IRAC bands by SMUVS.
\label{fig:colors}}
\end{figure*}

The colors of IRAC sources as measured by SMUVS are also consistent with what has been seen 
in other surveys.  In Fig.~\ref{fig:colors} we plot the [3.6]$-$[4.5] color of all SMUVS sources detected
in both IRAC bands and having 3.6\,$\mu$m magnitudes between 22 and 25.5\,AB mag.  The behavior
of these color distributions is essentially identical to what has been seen before by, e.g., S-CANDELS 
(Ashby et al.\ 2015), in particular it produces both the bimodal color distribution typically seen for 
relatively bright sources and the trend toward a redder, single-mode distribution at fainter fluxes.

The 14th and last columns of the three SMUVS catalogs contain data quality flags for the 
3.6 and 4.5\,$\mu$m photometry, respectively.  A flag of zero indicates no known issues with the photometry.   
Photometry for some sources was {\sl potentially} 
corrupted by the large halos around bright stars, which may have compromised the ability of {\tt StarFinder} 
to reliably estimate the local backgrounds.  
Sources potentially affected by background contamination have been assigned a data quality flag of 1.  
All cataloged sources having only single-band detections are also flagged.  A flag of 2 indicates a 
bright ($>23.0$\,mag) single-band detection.  Given the well-documented color distribution of IRAC-detected sources, 
which is reproduced by SMUVS (Fig.~\ref{fig:colors}), it is implausible that such bright sources 
(if real) would be absent in the off-band.  Instead, as we have verified by inspection, such detections are
artifacts of the shredding of extended sources.   A flag of 3 indicates a different single-band detection, 
i.e., a  faint ($<23.0$\,mag) one.  Unlike the bright single-band detections, faint sources {\sl can} be real, 
and arise when red or blue IRAC colors push the off-band flux below the SMUVS detection threshold.  
It is apparent by inspection, many such single-band detections are indeed apparent in the off-band, but 
at such low signal-to-noise ratios that they are not formally detected by {\tt StarFinder}.  

\subsection{Limitations of the SMUVS Catalogs}

\begin{figure*}[ht!]
\figurenum{9}
\epsscale{1.1}
\plotone{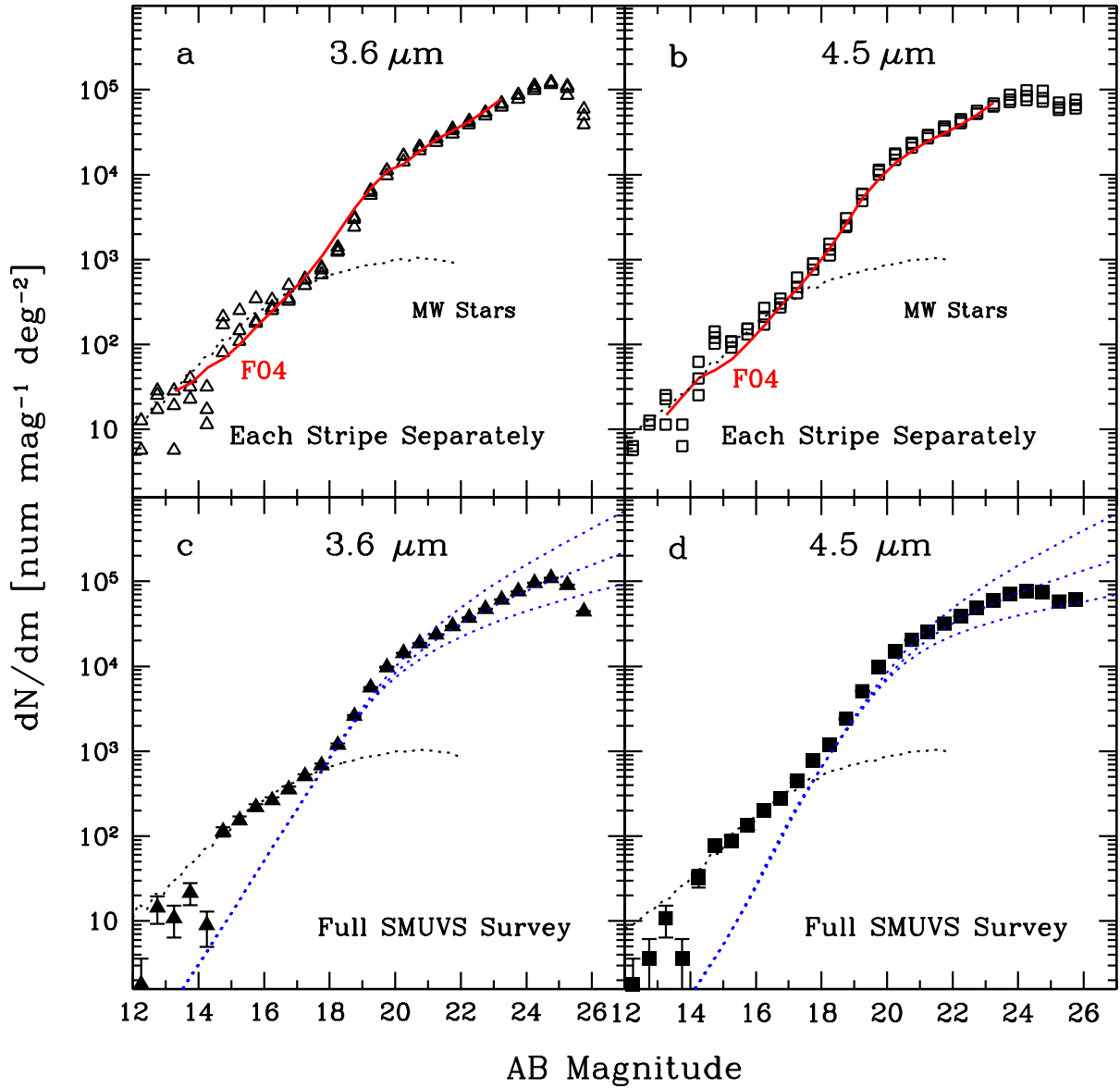}
\caption{Differential source counts in the COSMOS field measured in both warm IRAC bands by SMUVS.
Open symbols show the counts for individual stripes, while solid symbols indicate the counts summed 
over all three SMUVS stripes.  Error bars represent only the Poisson statistics arising from the numbers of galaxies in each magnitude bin.  The red lines in panels (a) and (b) indicate the incompleteness-corrected counts measured in the EGS by Fazio et al.\ (2004b). The dotted lines indicate estimated counts arising from Milky Way stars in COSMOS specifically, based on the DIRBE Faint Source Model at 3.5 and 4.9\,$\mu$m (Arendt et al.\ 1998; Wainscoat et al.\ 1992; Cohen 1993, 1994, 1995).  The blue dotted lines in the lower panels indicate the Helgason et al.\ (2012) model counts for those bands.
In both cases the IRAC counts closely follow the middle trend until rather faint magnitudes even in the absence of
completeness corrections, which have not been applied.
\label{fig:counts}}
\end{figure*}

The SMUVS catalog is not optimal for extended sources.  Such objects are likely to be 'shredded' into
multiple objects by the iterated PSF-fitting procedure.  Users should be cautious about SMUVS photometry 
of bright, extended sources.  As noted above, shredded sources can appear in the SMUVS catalogs as 
single-band detections, because extended sources are unlikely to be modeled by {\tt StarFinder} with spatially 
registered point sources in both bands.
We have flagged high-SNR but unmatched sources in the catalog so indicate that they likely arise from 
shredded objects.  Other unmatched sources are undoubtedly real, as corroborated by how closely 
the SMUVS source counts follow the deeper completeness-corrected counts from S-CANDELS 
(Fig.~\ref{fig:counts}), down to 25\,AB mag; these objects are  'lost'
from the off-band because of their colors.  Users of the SMUVS catalogs are cautioned to make use of the data
quality flags and to take the proximity of bright sources into account when interpreting the photometry.

Sources fainter than about 23\,AB mag will not be impacted by shredding, but brighter sources
could be if they are extended.  An attempt has been made to correct the SMUVS photometry 
in a statistical sense for modestly 
extended sources, following Ashby et al.\ (2013a), but this approach will be inadequate
for sources that are broader than 1--2$\times$ the IRAC FWHM.  
Users can examine the curve of growth through the SMUVS apertures as a 
means of verifying the photometry for individual sources.  
Well-characterized sources will have aperture magnitudes that agree
with each other, and with the PSF-fitted photometry.

The SMUVS catalog is not the best source of photometry for especially bright sources even if they
are pointlike.  To most efficiently photometer the faint, distant galaxies that are the primary objectives
for SMUVS, it was necessary to adopt a length scale on which to model the variations seen the backgrounds
of the SMUVS mosaics.  The length scale chosen -- 72$\times$ the FWHM of the PSFs used for photometry --
was a compromise between the competing needs to accurately fit both bright stars' outskirts and background
variations.  As a result, the magnitudes for Milky Way stars brighter than $\sim$13\,AB mag are systematically 
underestimated to varying degrees.  

The SMUVS counts are shown in Fig.~\ref{fig:counts} and Table~\ref{tab:counts}.  They appear 
broadly consistent with number counts 
based on UltraVISTA $HK_s$ priors (Deshmukh et al. (2017) down to $\sim$24.5\,mag, at which point 
incompleteness begins to have an impact.  At bright magnitudes the SMUVS counts
closely follow the Milky Way star counts model derived from DIRBE observations toward
 COSMOS (Arendt et al.\ 1998), except for very bright sources which are impacted by small
number statistics.   The dashed lines shown in Fig.~\ref{fig:counts} are not fits to the counts.  At fainter magnitudes
the counts closely follow the so-called 'default' model from Helgason et al.\ (2012).  The SMUVS counts follow
a linear trend all the way from the 'knee' of the counts at 20\,AB mag to faint count levels, where they depart
from the 'default' model at roughly [3.6]=25 and [4.5]=24.5\,AB mag.  We infer that the SMUVS counts begin
to suffer from significant incompleteness at about these levels.   By comparison, the not-completeness-corrected
counts shown for COSMOS in Figs.\ 17c and 18c of Ashby et al.\ (2015) depart from this trend at brighter magnitudes.
It therefore appears that the SMUVS catalogs are complete to significantly fainter levels than the earlier catalogs
built for the COSMOS field.  

Finally, users can refer to the IRAC color distributions in Fig.~\ref{fig:colors} as indicators of valid photometry.
Most (not all) sources with valid photometry will have IRAC colors in the range $-0.5<[3.6]-[4.5]<0.5$.  
If a cataloged SMUVS object has an IRAC color with an absolute value greater than unity, the underlying IRAC photometry should be treated with caution.

\begin{deluxetable}{l@{\hspace*{3em}}c@{\hspace*{3em}}c@{\hspace*{3em}}c@{\hspace*{3em}}c}
\tabletypesize{\footnotesize}
\tablewidth{344pt}
\tablecaption{Deep Source Counts in the IRAC Bands \label{tab:counts}}
\tablehead{
{Mag} & \multicolumn{2}{c}{3.6\,$\mu$m} & \multicolumn{2}{c}{4.5\,$\mu$m} \\
(AB) &  Counts & Unc & Counts & {Unc}
}
\startdata
13.25 &  1.032 & 0.188 &  1.032 & 0.188 \\
13.75 &  1.333 & 0.129 &  0.555 & 0.383 \\
14.25 &  0.953 & 0.209 &  1.509 & 0.104 \\
14.75 &  2.053 & 0.055 &  1.887 & 0.067 \\
15.25 &  2.188 & 0.047 &  1.944 & 0.062 \\
15.75 &  2.340 & 0.039 &  2.123 & 0.051 \\
16.25 &  2.421 & 0.036 &  2.299 & 0.041 \\
16.75 &  2.553 & 0.031 &  2.450 & 0.035 \\
17.25 &  2.704 & 0.026 &  2.657 & 0.027 \\
17.75 &  2.831 & 0.022 &  2.887 & 0.021 \\
18.25 &  3.073 & 0.017 &  3.079 & 0.017 \\
18.75 &  3.413 & 0.011 &  3.382 & 0.012 \\
19.25 &  3.749 & 0.008 &  3.705 & 0.008 \\
19.75 &  3.989 & 0.006 &  3.994 & 0.006 \\
20.25 &  4.157 & 0.005 &  4.179 & 0.005 \\
20.75 &  4.272 & 0.004 &  4.309 & 0.004 \\
21.25 &  4.372 & 0.004 &  4.410 & 0.004 \\
21.75 &  4.471 & 0.003 &  4.504 & 0.003 \\
22.25 &  4.573 & 0.003 &  4.589 & 0.003 \\
22.75 &  4.676 & 0.003 &  4.689 & 0.003 \\
23.25 &  4.784 & 0.002 &  4.776 & 0.002 \\
23.75 &  4.881 & 0.002 &  4.851 & 0.002 \\
24.25 &  4.982 & 0.002 &  4.892 & 0.002 \\
24.75 &  5.038 & 0.002 &  4.877 & 0.002 \\
25.25 &  4.959 & 0.002 &  4.760 & 0.002 \\
25.75 &  4.647 & 0.003 &  4.793 & 0.002 \\
\enddata
\tablecomments{Differential number counts in the COSMOS field as measured in the three SMUVS stripes
in both operable IRAC bands, expressed in terms of log$(N)$\,mag$^{-1}$\,deg$^{-2}$.  
Uncertainties are 1$\sigma$ estimates based solely on
the number of sources in each bin, and do not reflect calibration errors, systematic effects or incompleteness corrections, which were not applied.}
\end{deluxetable}


\acknowledgments

This work is based on observations made with the {\sl Spitzer Space Telescope}, 
which is operated by the Jet Propulsion Laboratory, California Institute of Technology 
under a contract with NASA.  Support for this work was provided by NASA through an 
award issued by JPL/Caltech.  The Cosmic Dawn Center is funded by the DNRF.

\vspace{5mm}
\facilities{Spitzer(IRAC)}





\allauthors

\listofchanges

\end{document}